\documentclass[10pt,
aps,
prl,
twocolum,
longbibliography
]{revtex4-2}
\usepackage{graphicx}
\usepackage{amsmath,amssymb,bm}
\usepackage[utf8]{inputenc}
\usepackage{comment}
\usepackage[svgnames]{xcolor}
\usepackage{siunitx}
\usepackage[colorlinks,linkcolor=MediumBlue,citecolor=MediumBlue,urlcolor = MediumBlue]{hyperref}

\renewcommand{\selectlanguage}[1]{}

\newcommand{\vect}[1]{\boldsymbol{#1}}
\newcommand{\unitvect}[1]{\hat{\vect{#1}}}
\newcommand{\mean}[1]{\left \langle {#1} \right \rangle}
\newcommand{\abs}[1]{\left| {#1} \right|}
\newcommand{\dd}[1]{\text{d}{#1}}
\newcommand{\current}{\langle \dot{x} \rangle}
\newcommand{\order}[1]{\mathcal{O}({#1})}

\newcommand{\xresetleft}{x_\mathrm{l}}
\newcommand{\xresetright}{x_\mathrm{r}}
\newcommand{\Tresetleft}{T_\mathrm{l}}
\newcommand{\Tresetright}{T_\mathrm{r}}

\newcommand{\vmax}{v_\alpha}
\newcommand{\Vscale}{\mean{v}_T}
\def\CABP{cABP}
\def\fmin{F_{\rm min}}
\def\fmax{F_{\rm max}}
\newcommand{\thetaconemin}{\theta_c} 
\newcommand{\dt}{\Delta t}

\DeclareMathOperator\artanh{artanh}

\makeatletter
\renewcommand{\p@section}{Appendix~\thesection} 
\renewcommand{\p@figure}{Fig.~} 
\makeatother

\usepackage{titlesec}
\titleformat{\paragraph}[runin]{\normalsize\itshape}{\theparagraph}{}{}[---\hspace*{-6pt}]

\date{}

\begin{document}
\title{Moving against the odds with a cyclic active Brownian ratchet:\\ current enhancement and reversal}
\author{Theo Spornhauer$^{1,2}$}
\email{theo.spornhauer@ds.mpg.de}
\author{Beno\^it Mahault$^{3,1}$}
\email{benoit.mahault@umontpellier.fr}
\author{Johannes Zierenberg$^{1,2}$}
\email{johannes.zierenberg@ds.mpg.de}
\affiliation{
  \mbox{$^1$ Max Planck Institute for Dynamics and Self-Organization, Am Fassberg 17, 37077 G{\"o}ttingen, Germany}\\
  \mbox{$^2$ Institute for the Dynamics of Complex Systems, University of G\"ottingen, G\"ottingen, Germany}\\
  \mbox{$^3$ Laboratoire Charles Coulomb (L2C), Universit{\'e} de Montpellier, CNRS, Montpellier, France}
}

\date{\today}
\begin{abstract}
    We show that periodically modulating the self-propulsion of active particles in an asymmetric potential can both enhance rectified transport and reverse its direction, even when the propulsion speed itself never changes sign.
    Current enhancement results from the nonlinear response of active transport to propulsion strength, whereas current reversal emerges from the interplay between activity-induced spreading and passive relaxation.
    Our theoretical analysis identifies the regimes governed by these distinct rectification mechanisms and predicts the finite range of driving periods over which current reversal occurs.
    These results establish cyclic activity as a means of dynamically controlling rectified active transport, with potential applications to the selective sorting of active particles.
\end{abstract}
\maketitle

\twocolumngrid

\paragraph{Introduction}
Biological agents are intrinsically exposed to fluctuating environments, where dynamic cues induce time-dependent changes in their activity.
A prominent example is the diurnal cycle, which drives periodic variations in the motility and metabolism of photosynthetic microorganisms such as cyanobacteria~\cite{cohen_circadian_2015, wilde_light-controlled_2017} and \textit{Chlamydomonas}~\cite{jin_diurnal_2020}, as well as the diel vertical migration of phytoplankton~\cite{hays_review_2003, morales_individual_2026}. 
More broadly, cyclic activity is widespread across biological scales: many organisms alternate between foraging, feeding, and resting, while human populations exhibit daily and weekly variations in mobility and contact rates that shape the spread of infectious diseases~\cite{earn_simple_2000, zierenberg_how_2023}. 
Cyclic activity can also emerge spontaneously through collective synchronization, for instance as pulsatile mechanical activity or oscillatory chemical signaling~\cite{schillers_real-time_2010, gregor_onset_2010}. 
Beyond living systems, externally imposed cyclic activity offers promising routes for controlling artificial active matter, 
for example through light-responsive propulsion~\cite{rey_light_2023} as well as actuated colloidal particles by electric~\cite{bricard_emergence_2013,yan_reconfiguring_2016,zhang_guiding_2022}, or acoustic~\cite{mcneill_acoustically_2023} fields.

Time-dependent activity has been studied primarily in homogeneous environments. 
For example, periodic swimming strokes can enhance long-time diffusion even when they produce no net displacement~\cite{lauga_enhanced_2011}, while stochastic switching between motile and nonmotile states can give rise to anomalous diffusion and non-Gaussian transport statistics~\cite{doerries_apparent_2022,datta_random_2024,santra_dynamics_2024}.
Propulsion protocols with zero time-averaged velocity, and hence periodic reversals of the propulsion direction, have also been shown to generate current reversals in active ratchets~\cite{benjamin_transport_2026} and suppress motility-induced phase separation~\cite{kailasham_effect_2023,cates_motility-induced_2015}. 
Yet, how cyclic modulations of activity affect the transport of active particles in structured environments remains largely unexplored.

In this work, we investigate the transport of a cyclic active Brownian particle (\CABP), 
defined as an ABP~\cite{romanczuk_active_2012, bechinger_active_2016} whose self-propulsion speed $v(t)$ varies periodically, 
in an asymmetric potential which provides a paradigmatic model of an active ratchet~\cite{reichhardt_ratchet_2017}. 
For a stepwise protocol alternating between active and passive phases, we show that cyclic activity can enhance the ratchet current and reverse its direction even in the absence of imposed self-propulsion reversals. 
Our theoretical analysis elucidates the mechanisms underlying these effects and quantitatively predicts the parameter regime in which current reversal occurs. 
These results demonstrate that cyclic activity can strongly modify transport in active ratchets, 
offering new perspectives for the dynamical control of rectified active motion.

\begin{figure}[t]
    \includegraphics[width=\linewidth]{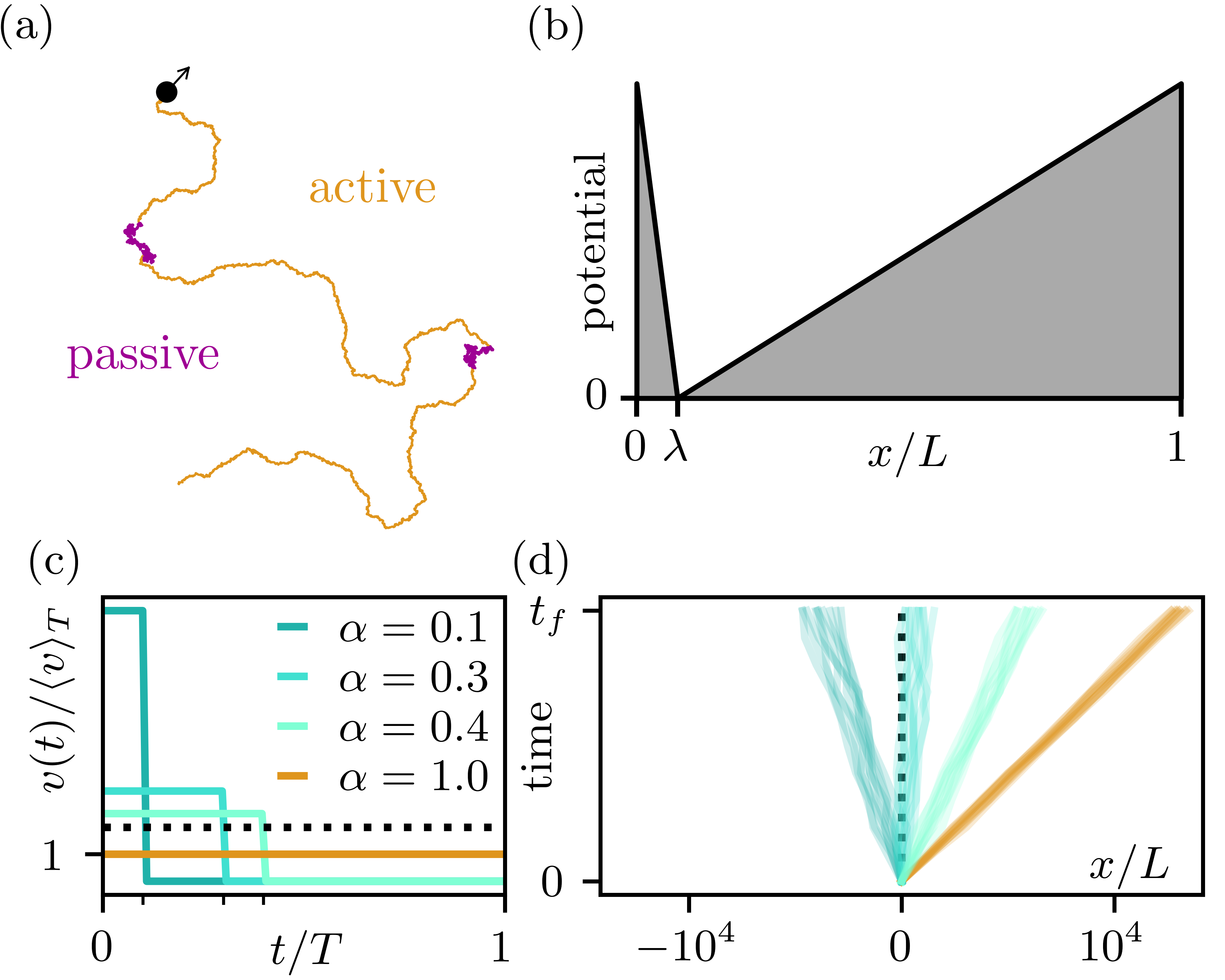}
    \caption{
    Quasi-one-dimensional cyclic active ratchet.
    (a) Example trajectory of a free \CABP\ alternating between active and passive states. 
    (b) Asymmetric, piecewise linear ratchet potential, periodic in $x$ and infinitely extended along $y$.
    (c) Time series of the cyclic self-propulsion speed over a single period for different active fractions $\alpha$.    
    The dotted line indicates $\fmax = \lambda^{-1}$, the force of the steep ratchet side. 
    (d) Individual trajectories corresponding to (c).
    For each value of $\alpha$, 10 trajectories are displayed.
    Decreasing $\alpha$, cyclic activity can reverse the steady-state current direction.
    Parameters: $\Vscale=10$, $T=2\pi10^{-2}$. In (d), $t_f = 5000$.
    }
    \label{fig:1}
\end{figure}

\paragraph{Cyclically driven active ratchet} 
We consider an overdamped \CABP\ moving in two dimensions within an external potential $V(\vect{x})$. 
The particle position $\vect{x}$ and orientation $\theta$ obey the Langevin equations 
\begin{align} \label{eq_Langevin} 
\dot{\vect{x}} &= v(t)\unitvect{e}(\theta)-\mu\nabla V(\vect{x}), 
& \dot{\theta} &= \sqrt{2D_r}\,\eta, 
\end{align} 
where $\unitvect{e}(\theta)=(\cos\theta,\sin\theta)$ is the self-propulsion direction and $\eta$ is Gaussian white noise with unit variance. 
The parameters $\mu$ and $D_r$ denote the particle mobility and rotational diffusivity, respectively. 
Except for the self-propulsion speed $v(t)$, all coefficients in Eq.~\eqref{eq_Langevin} are time-independent~\footnote{We have verified that adding weak translational diffusion to Eq.~\eqref{eq_Langevin} does not affect our results.}.

For simplicity, we consider a stepwise cyclic speed protocol with period $T$, 
alternating between active and passive phases with speeds $\vmax$ and $0$, respectively: 
\begin{equation} \label{eq:velocity} 
v(t)= \begin{cases} 
	\vmax, & t\in[0,\alpha T),\\ 
	0, & t\in[\alpha T,T), 
	\end{cases}
\end{equation} 
where $\alpha\in(0,1]$ is the active fraction of each cycle. 
The time-averaged speed is therefore $\Vscale=\alpha\vmax$. 
In free space, this protocol produces alternating persistent running and diffusive phases, 
as illustrated in \ref{fig:1}(a). 
Unlike run-and-tumble motion~\cite{solon_active_2015}, the state-switches induced by Eq.~\eqref{eq:velocity} are deterministic, 
while the rotational diffusivity remains constant over time.

The \CABP\ dynamics defined by Eqs.~\eqref{eq_Langevin} and~\eqref{eq:velocity} could be realized using experimental platforms that allow for dynamical motility actuation. 
Examples include light-responsive active colloids~\cite{jiang_active_2010, buttinoni_active_2012, palacci_living_2013, bregulla_stochastic_2014, gomez-solano_tuning_2017},
bacteria~\cite{walter_light-powering_2007, arlt_painting_2018, frangipane_dynamic_2018, koumakis_dynamic_2019, bianchi_dynamic_2026}, and robots~\cite{mijalkov_engineering_2016}, 
as well as active colloids actuated by electric~\cite{bricard_emergence_2013,yan_reconfiguring_2016,zhang_guiding_2022} or acoustic fields~\cite{mcneill_acoustically_2023}.

Following the ratchet principle~\cite{reimann_brownian_2002, metzger_revisiting_2026}, 
we consider a one-dimensional sawtooth potential $V(x)$ of period $L$ and asymmetry parameter $\lambda\in(0,\tfrac{1}{2}]$, 
as shown in~\ref{fig:1}(b). 
Within an elementary cell $x\in[0,L]$, the potential has a minimum at $x=\lambda L$ 
and is characterized by a narrow region with strong confining force $\fmax = V_0/(\lambda L)$, 
together with a wider region with weaker force $\fmin = V_0/[L(\lambda - 1)]$. 

We measure length, time, and energy in units of $L$, $D_r^{-1}$, and $V_0$, respectively, such that $L=D_r=V_0=1$.
The remaining dimensionless mobility is fixed to $\mu=1$.  
For $\lambda=0.05$, a particle propelled at constant speed, corresponding to $\alpha=1$, exhibits standard active rectification~\cite{angelani_active_2011, ghosh_self-propelled_2013, reichhardt_ratchet_2017}.
When oriented along the $x$ direction, it climbs each potential side at an effective speed reduced by the relevant opposing force-induced drift.
The particle is thus less likely to cross the barrier on the steep side, thereby generating a net current toward the shallow side.
In particular, for $\vmax \in [\abs{\fmin},\fmax]\approx[1.05,20]$, the particle can climb the shallow side but not the steep one, while rectification is strongest for $\vmax\approx\fmax$.

\begin{figure}
    \centering
    \includegraphics[width=\linewidth]{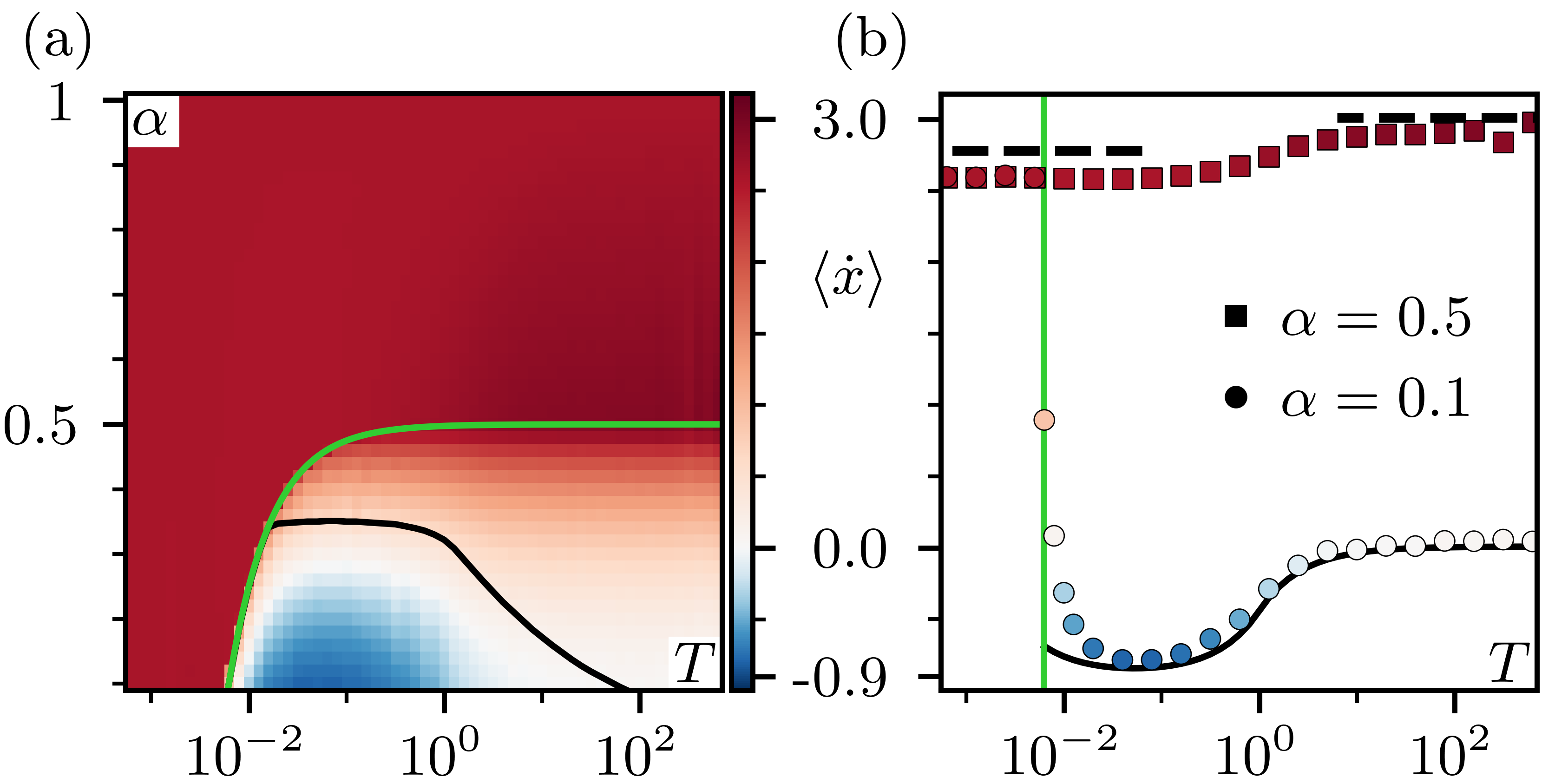}
    \includegraphics[width=\linewidth]{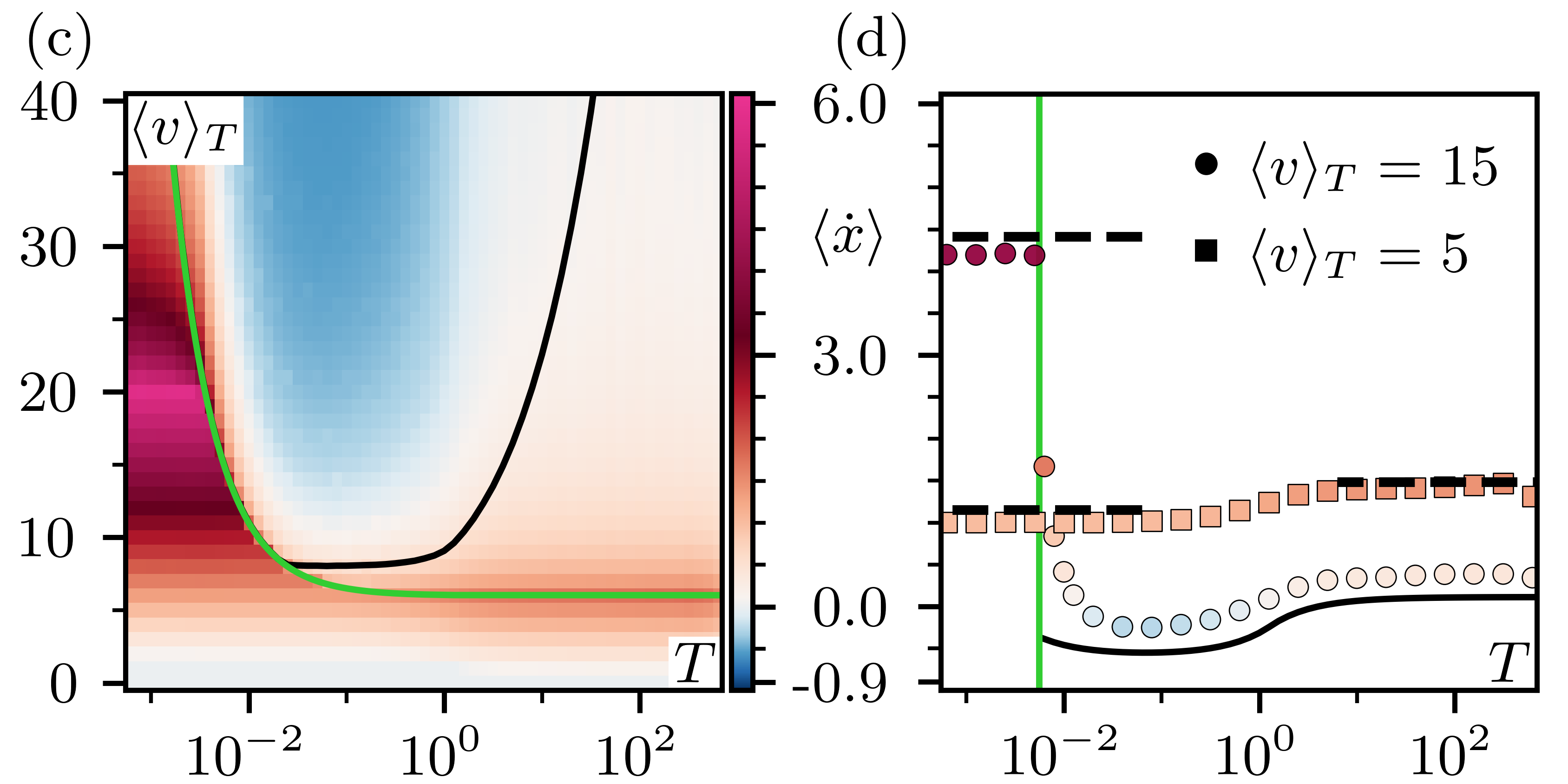}
    \caption{
    Steady-state average current of \CABP s in an asymmetric ratchet potential.
    (a) Phase diagram in the ($T,\alpha$) plane highlighting regions of positive (red), zero (white) and negative (blue) currents measured from numerical simulations for fixed $\Vscale = 10$.
    The green line indicates the boundary given in Eq.~\eqref{eq_omega_plus}, while the black line marks the analytical estimate of the transition to negative current.
    (b) Steady-state current as a function of the driving period for several values of $\alpha$.
    Simulation data is displayed with the colored symbols, while our analytical estimate of the negative current is shown with the continuous black line.
    Approximations of the current in the limiting regimes of zero and infinite period [Eq.~\eqref{eq_J_steady}, see text] are shown with the horizontal black dashed lines.
    (c,d) Same as (a,b) for fixed $\alpha = 0.3$ and varying mean speed $\Vscale$.
    }
    \label{fig:2}
\end{figure}

Cyclic activity can drastically alter this rectification mechanism, as illustrated in~\ref{fig:1}(c,d). 
Fixing $\Vscale=10$ well within the active-rectification regime,
the ratchet exhibits a positive steady-state current for constant speed ($\alpha=1$) [orange lines in~\ref{fig:1}(c,d)]. 
Reducing $\alpha$ at fixed $\Vscale$ increases the active speed, $\vmax=\Vscale/\alpha$. 
Once $\vmax>\fmax$, the particle can surmount both edges of the potential during the active phase, weakening the rectification and reducing the current [light green lines in~\ref{fig:1}(c,d)]. 
Remarkably, for certain driving frequencies, further decreasing $\alpha$ \emph{reverses the current}: the particle then crosses the steep barrier more frequently than the shallow one [dark turquoise lines in~\ref{fig:1}(c,d)].

\paragraph{Phase behavior}
We characterize the steady-state current \mbox{$\current=\lim_{t\to\infty} \langle x(t)\rangle / t$} as a function of the active fraction $\alpha$, the driving period $T$, and the average self-propulsion speed $\Vscale$ in~\ref{fig:2}. 
We restrict our analysis to the case $\vmax>\abs{\fmin}$, such that the particle is not pinned at the potential minimum. 
The phase diagrams in the $(T,\alpha)$ [\ref{fig:2}(a)] and $(T,\Vscale)$ [\ref{fig:2}(c)] planes, 
obtained from numerical simulations of Eq.~\eqref{eq_Langevin}, 
reveal distinct regimes of positive (red), vanishing (white), and negative (blue) current.
Positive currents primarily occur when the active phase is too short for a particle to cross the steep side of the potential~\footnote{For our parameters, except at very short periods or for $\alpha\approx1$, a particle located on the steep side at the beginning of the passive phase fully relaxes to the potential minimum before the next active phase.}. 
Neglecting rotational diffusion, we approximate a bound for this regime by requiring a particle with constant speed $-\vmax + \fmax$ to travel a distance $\lambda$ within time $\alpha T_+$, yielding 
\begin{align} \label{eq_omega_plus} 
	\left(\Vscale\lambda-\alpha\right)T_+=\lambda^2, 
\end{align} 
presented as green lines in \ref{fig:2}. 

\begin{figure}
    \centering
    \includegraphics[width=\linewidth]{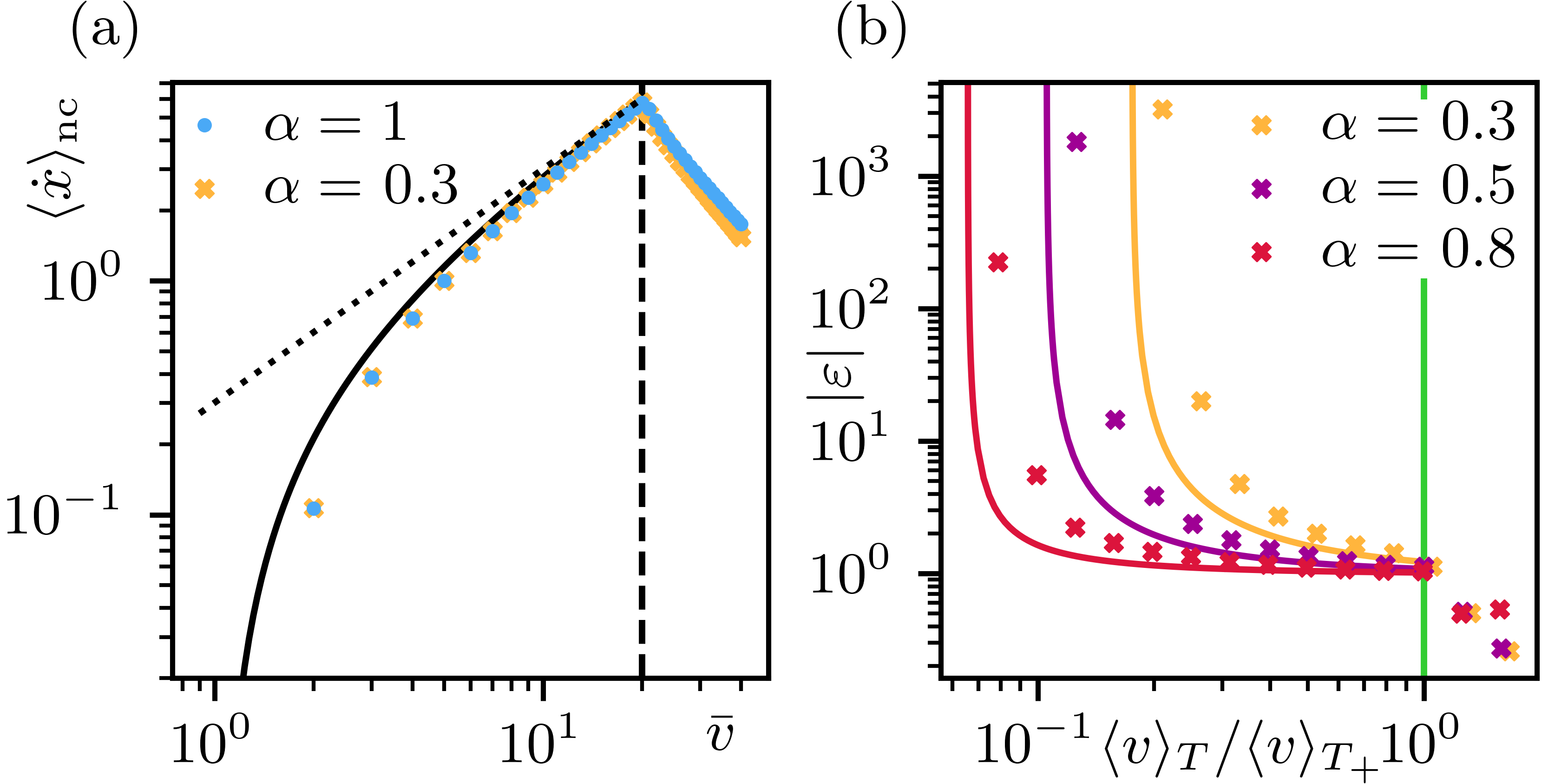}
    \caption{
    Current enhancement by cyclic activity.
    (a) Steady-state current as a function of the self-propulsion speed $\bar v$.
    Blue and orange markers show data from numerical simulations for the non-cyclic and self-averaging cyclic case ($T=2\pi10^{-4}$ with $\bar v = \Vscale$), respectively.
    The black lines display the prediction of Eq.~\eqref{eq_J_steady} (solid) and  the bound $\bar v = \fmax$ (dashed). 
    The dotted line, with slope~1, is included as a guide to the eye.
    (b) Current enhancement ratio $\varepsilon$ as a function of the normalized mean speed.
    Solid lines show the prediction from Eq.~\eqref{eq:current_enhancement_epsilon} 
    and data points correspond to numerical simulations.
    The numerical data points in (b) use the current for periods $T=2 \pi 10^{-4}$ and $T=2 \pi 10^{2}$ to approximate $\varepsilon$.
    }
    \label{fig:3}
\end{figure}

In the fast driving limit $T \to 0$, the \CABP\ effectively behaves as a standard ABP with constant speed $\Vscale$. 
The ratchet current in the limit $T \to 0$ can thus be inferred from its counterpart in the non-cyclic case 
with constant speed $\bar v = \Vscale$. 
For sufficiently large speeds $\bar v\in(|\fmin|,\fmax)$, we obtain the approximated expression
\begin{align} 
\label{eq_J_steady}
    \current_{\rm nc} & = \frac{\sqrt{\bar v^2- (1-\lambda)^{-2}}}{\pi}\left( 1 + \frac{\lambda^2}{2} \right) + \mathcal{G}(\lambda, \bar v) ,
\end{align}
where the derivation and the function $\mathcal{G}$ are given in \hyperref[app:non_cyclic_current]{Appendix A}. 
As shown by the black dashed curves in~\ref{fig:2}(b,d), and by the solid curve in~\ref{fig:3}(a), 
Eq.~\eqref{eq_J_steady} evaluated at $\bar v=\Vscale$ provides a reasonable, parameter-free estimate of the current in the limit $T\to0$.

\paragraph{Current enhancement}
Interestingly, we observe in \ref{fig:2}(b,d) that the current systematically increases as $T \to \infty$.
In this limit, the system reaches its steady state on time scales much shorter than the driving period, such that its long-time evolution can be considered as adiabatic.
With a negligible current from the passive phase, we thus expect for $T\to \infty$ that $\current\to\alpha \current_{\rm nc}$,
where the latter is evaluated at $\bar v = \vmax$.

\ref{fig:3}(a) shows that $\current_{\rm nc}$ grows faster than linearly with $\bar v$ for $\bar v \in [\abs{\fmin}, \fmax]$. 
Since $\vmax = \Vscale/\alpha \geq \Vscale$, concentrating the same mean propulsion into increasingly strong active phases enhances the cycle-averaged current. 
We quantify this enhancement through the ratio
\begin{equation}
    \label{eq:current_enhancement_epsilon}
    \varepsilon =
    \frac{\lim_{T \to \infty}\current}
    {\lim_{T \to 0}\current} 
    \simeq
    \frac{\alpha \current_{\rm nc}|_{\bar v = \vmax}}
    {\current_{\rm nc}|_{\bar v = \Vscale}} ,
\end{equation}
while the superlinear dependence of $\current_{\rm nc}$ on speed implies that $\varepsilon>1$ when active rectification operates.
\ref{fig:3}(b) further shows that the enhancement factor~\eqref{eq:current_enhancement_epsilon} is largest as $\Vscale \to \abs{\fmin}$ and formally diverges at $\Vscale=|\fmin|$, where the steadily propelled particle becomes pinned by the potential.

\paragraph{Current reversal}
As described above, for $T>T_+$, an active particle with suitable orientation can escape from a potential minimum along the steep side within a single active phase. 
For slow periodic driving ($T \gg T_+$), active rectification consequently weakens, resulting in an almost vanishing net current. 
At intermediate periods centered around $T=\mathcal{O}(10^{-1})$, however, \ref{fig:2} reveals a regime of negative current.

\begin{figure}
    \centering
    \includegraphics[width=\linewidth]{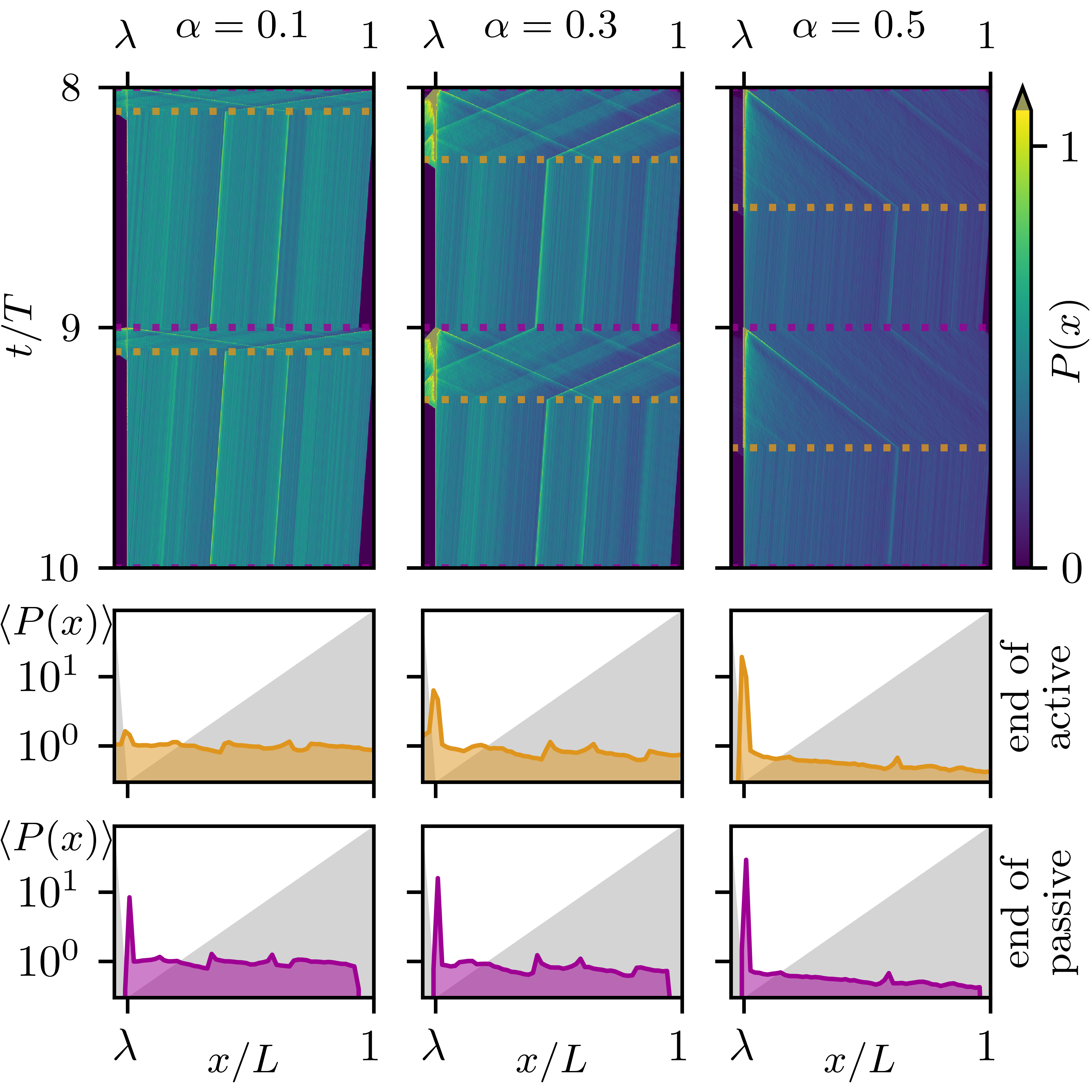}
    \caption{Probability density function of the particle's $x$-position in a unit-cell during the cyclic activity with period $T=2 \pi 10^{-2}$ and active fractions $\alpha=0.1, 0.3, 0.5$ from left to right.
    (Top row) Time evolution over two steady-state periods is displayed as kymograph plots.
    (Middle row) Probability density at the end of the active phase averaged over five periods.
    (Bottom row) Probability density at the end of the passive phase averaged over five periods.}
    \label{fig:4}
\end{figure}

To elucidate the origin of such current reversal, we show in \ref{fig:4} kymographs of the particle density in the negative current regime over several driving periods for increasing values of the active fraction $\alpha$. 
At the beginning of each active phase, the density is strongly peaked at the potential minimum $x=\lambda$. 
For $\vmax>\fmax$, switching on activity generates two density fronts that originate from the minimum and propagate in opposite directions (top row of \ref{fig:4}). 
When self-propulsion is much larger than the drift induced by the potential, these fronts travel nearly equal distances during the time $\alpha T$, thus highlighting the absence of net displacement over the active phase.
Since the traveling fronts are constantly generated and disperse over timescales $\simeq 1/D_r$, corresponding to a few periods in the regime of interest, 
the particle density is nearly uniform at the end of the active phase, although a residual peak at $x=\lambda$ may persist for weaker self-propulsion (middle row of \ref{fig:4}).

When activity is switched off, the particle relaxes towards the accessible potential minimum.
Because the shallow side---which drives particles to the left---occupies a larger fraction of space than the steep side [\ref{fig:1}(b)], this passive dynamics generates a net negative current (middle and bottom rows of \ref{fig:4}).
Increasing $\alpha$ reduces $\vmax$ and thereby restores active rectification.
Moreover, it also leaves the particles more strongly localized around the potential minimum at the end of the active phase, thus reducing the negative displacement in the subsequent passive phase.
As a consequence, the asymmetry of the potential once again favors transport in the positive direction.
Current reversals therefore result from the combination of two factors: (i) the nearly zero net displacement during the active phase, caused by the high particle velocity and the homogeneous distribution of particle orientations, and (ii) the nearly uniform density profile at the end of the active phase, such that rectification effectively operates solely during the passive phase.

We now use these observations to estimate the parameter range over which current reversal occurs.
The negative current is observed for driving periods typically much smaller than the rotational diffusion time $1/D_r$.
We therefore consider a simplified scenario in which, during the active phase, a particle moves with constant self-propulsion $\vmax > \fmax$ along either $\pm \hat{\bm x}$ without changing its orientation.
We estimate its average velocity during the active phase from the times $\tau_\pm$ required to traverse one spatial period of the potential while moving along $\pm \hat{\bm x}$.
These times are given by (see \hyperref[app:displacement_current]{Appendix B} for calculation details)
\begin{equation} \label{eq_taupm}
	\tau_\pm =
	\frac{1}{\vmax} \left[ 1 - \frac{\lambda}{1 \pm \lambda \vmax} - \frac{1-\lambda}{1 \mp (1-\lambda)\vmax} \right].
\end{equation}
Using $\tau_\pm^{-1}$ to approximate the effective particle velocity (recall that $L=1$ in our units), 
we estimate the active contribution to the mean current as
$\current_{\rm a} = \alpha \left( \tau_+^{-1} - \tau_-^{-1} \right)$.
It is straightforward to show that $\current_\mathrm{a} > 0$, confirming that the active dynamics systematically rectifies motion along $+\hat{\vect{x}}$.
In the large $\vmax$ limit, however, the average particle velocity becomes essentially independent of its orientation, 
such that $\current_\mathrm{a}\to0$.

To estimate the passive contribution, 
we assume that the particle position is uniformly distributed at the beginning of the passive phase and then drifts toward the potential minima.
As detailed in \hyperref[app:displacement_current]{Appendix B}, 
when $(1-\alpha)T > (1-\lambda)^2$, the particle reaches a minimum during the passive phase regardless of its initial position, yielding the mean current $\current_{\rm p} = - \tfrac{1}{T}\left( \tfrac{1}{2} - \lambda \right)$.
For shorter periods relevant to the negative-current regime, a particle initially located near the tip of the shallow side may not reach the minimum within the duration of the passive phase.
In this case, the passive contribution becomes
\begin{equation} \label{eq_passive_current_calc}
    \current_\mathrm{p} = - \frac{1}{T} \left( \frac{1}{2} - \lambda \right) + \frac{1}{2T} \left( 1- \lambda - \frac{(1-\alpha)T}{1-\lambda} \right)^2.
\end{equation}
For shorter periods, a particle located on the steeper branch may also fail to reach the minimum (see \hyperref[app:displacement_current]{Appendix B} for details).
However, this regime typically corresponds to $T < T_+$ and therefore lies outside the parameter range relevant to our analysis.
The mean steady-state current is finally estimated as $\current = \current_{\rm a} + \current_{\rm p}$.

Black lines in \ref{fig:2}(a,c) correspond to the condition $\current = 0$ and delimit the parameter regions in which the theory predicts current reversal.
They show remarkable agreement with the numerical results.
Furthermore, \ref{fig:2}(b,d) shows that our analysis quantitatively captures the variations of $\current$ with the driving frequency without fitting parameters.
In particular, the predicted current reaches its minimum when the duration of the passive phase satisfies $(1-\alpha)T = \lambda(1-\lambda) = -\lambda/\fmin \approx 0.05$, 
which corresponds to the point where the absolute displacement generated along the shallow side during the passive phase equals the spatial extent $\lambda$ of the steep side.

\paragraph{Conclusion} We have shown that cyclic activity qualitatively alters transport in active ratchets, 
 producing both enhanced rectification and current reversals. 
 At fixed mean self-propulsion speed, cyclic activity can substantially enhance the ratchet current. 
 This enhancement is strongest when the mean propulsion approaches the drift induced by the shallow side of the potential, where an ABP with constant activity becomes pinned, whereas sufficiently strong activity bursts still generate directed motion. 
 Cyclic activity can also reverse the current through a mechanism distinct from conventional active rectification. 
 Strong activity bursts broaden and nearly homogenize the particle distribution, while the subsequent passive phase generates a net negative flux because the basin of attraction extends farther along the shallow side of the potential.
 
In the current-reversal regime, the cyclic active ratchet therefore operates analogously to a flashing or temperature ratchet~\cite{reimann_brownian_2002}. 
The active phase plays the role of an off-potential or high-temperature phase, during which the particle distribution broadens, 
while the passive phase corresponds to an on-potential or low-temperature phase, 
during which particles relax toward the potential minima, whose asymmetry induces rectification. 
As $\alpha$ and $T$ are varied, the \CABP\ thus interpolates between an active and flashing-like ratchet,
while the current reversal marks the crossover between regimes dominated by either of these two mechanisms.

Current reversals are a recurring feature of Brownian motors~\cite{reimann_brownian_2002, hanggi_artificial_2009}. 
They have been identified most extensively in rocking ratchets, where they may be induced by changes in the applied load, noise strength, particle inertia, or driving parameters~\cite{ajdari_rectified_1994,bartussek_periodically_1994, jung_regular_1996, reimann_quantum_1997}. For active particles in asymmetric potentials, spatial variations of the propulsion speed~\cite{pototsky_rectification_2013} and many-body effects~\cite{metzger_revisiting_2026} can likewise reverse the direction of transport. 
Here, in contrast, the reversal is driven entirely by the temporal structure of the activity, while both the potential and the mean propulsion speed remain fixed. 
Because particles with different propulsion speeds can exhibit different current directions over the same frequency window, suitably tuning the shape and frequency of the activity cycle could enable their separation.
Cyclic active ratchets may therefore complement existing geometric~\cite{reichhardt_ratchet_2017,volpe_microswimmers_2011} and chirality-based~\cite{mijalkov_sorting_2013} sorting strategies, with the distinctive advantage that their selectivity can be adjusted dynamically.

\begin{acknowledgments}
We acknowledge support from the Max Planck Society and, especially, the Department of Living Matter Physics, Ramin Golestanian and Viola Priesemann.
T.~S. thanks the Charles Coulomb laboratory for hosting him during part of this work.
J.~Z. was funded by the Deutsche Forschungsgemeinschaft (DFG) via EXC 2067/1-390729940 and SFB 1690/1, A07.
\end{acknowledgments}

\bibliography{active_ratchet}

\onecolumngrid

\clearpage

\vspace{12pt}
\noindent\hfill {\large \bf End Matter} \hfill
\vspace{12pt}

\twocolumngrid

\renewcommand \thefigure{A\arabic{figure}}
\setcounter{figure}{0}
\setcounter{equation}{0}
\renewcommand{\theequation}{A\arabic{equation}}

\paragraph{\label{app:non_cyclic_current}Appendix A: stationary current in a non-cyclic active ratchet}
In their dimensionless form and projecting the position dynamics over $\hat{\vect{x}}$, 
Eqs.~\eqref{eq_Langevin} with constant self-propulsion speed $\bar v$ read
    \begin{equation} \label{eq_Langevin_appA}
\dot{x} = \bar v \cos\theta + F(x), \qquad
\dot{\theta} = \sqrt{2} \eta,
    \end{equation}
where $F(x) = \fmax = \lambda^{-1}$ for $0 \le x < \lambda$, and $F(x) = \fmin = (\lambda-1)^{-1}$ for $\lambda \le x < 1$.
The corresponding Fokker-Planck equation for the probability distribution $P(x,\theta,t)$ 
is then given by
\begin{equation}\label{eq_FokkerPlanck_1D_appendix}
    \partial_t P + \partial_x \left[\bar v \cos \theta P + F(x) P \right] = \partial^2_{\theta\theta} P.
\end{equation}
Hereafter, we look for stationary solutions of Eq.~\eqref{eq_FokkerPlanck_1D_appendix}
in the limit of large self-propulsion velocity while imposing $\bar v \in [\abs{\fmin};\fmax]$,
such that $F(x) = \order{\bar v}$,
in order to retain the rectifying effect of the potential.
While the calculation can be straightforwardly expanded to include the case $\bar v > \fmax$, 
here we restrict the analysis to self-propulsion velocities where we observe current enhancement in the cyclic case.

Assuming the ansatz $P(x,\theta) = \sum_{k\geq 0} \bar v^{-k} P_k(x,\theta)$ and keeping terms of same order in $\bar v$ in Eq.~\eqref{eq_FokkerPlanck_1D_appendix}, we then obtain to leading order, $\order{\bar v}$,
\begin{equation}
    \partial_x \left\{ \left[\bar v \cos\theta + F(x) \right] P_0(x,\theta) \right\} = 0.
\end{equation}
Integrating over $x$, we have
\begin{equation}
    P_0(x,\theta) = \frac{j_0(\theta)}{\bar v \cos \theta + F(x)},
\end{equation}
where the integration constant $j_0(\theta)$ is analogous to a $\theta$-dependent density current.
Since the orientation dynamics in Eq.~\eqref{eq_Langevin_appA} only consists of rotational diffusion, 
we assume that the marginal $\mathcal{P}_0(\theta) = \int_0^1 \dd{x} P_0(x, \theta)$ is uniform, 
yielding $\mathcal{P}_0(\theta) = \tfrac{1}{2\pi}$ and therefore:
\begin{equation}
    \label{eq_j0_appendix}
     j_0(\theta) = \frac{1}{2\pi} \left[ \frac{\lambda}{\bar v \cos\theta + \fmax} 
     + \frac{1-\lambda}{\bar v \cos\theta + \fmin} \right]^{-1} .
\end{equation}
However, the particle can navigate in the potential---and hence, generate a nonzero current---when 
$\bar v \cos\theta > \abs{\fmin}$, meaning that $\abs{\theta} < \thetaconemin \equiv \arccos(\abs{\fmin}/\bar v)$.
Otherwise, the particle remains trapped in the potential minimum and $j_0(\theta)=0$.
Therefore the distribution reads
\begin{equation} \label{eq_P0_app}
    P_0(x, \theta) = \begin{cases}
        \frac{j_0(\theta)}{\bar v \cos\theta + F(x)} & \mbox{for} \abs{\theta} < \thetaconemin
        \\
        \frac{1}{2 \pi} \delta(x-\lambda) & \mbox{otherwise}
    \end{cases}, 
\end{equation}
with $j_0(\theta)$ given in Eq.~\eqref{eq_j0_appendix}
and where $\theta \in (-\pi;\pi]$.

To include leading order corrections to~\eqref{eq_P0_app} induced by rotational diffusion,
we write Eq.~\eqref{eq_FokkerPlanck_1D_appendix} retaining terms of $\order{1}$:
\begin{equation}
    \partial_x \left[ \left( \cos\theta + \frac{F(x)}{\bar v} \right) P_1 \right] = \partial^2_{\theta\theta} P_0,
\end{equation}
whose solution is
\begin{equation*}
    P_1(x,\theta) = \frac{\bar v}{\bar v \cos\theta + F(x)} \left[j_1(\theta) + \int_x \dd{x'} \partial^2_{\theta\theta} P_0(x', \theta) \right],
\end{equation*}
where $j_1(\theta)$ is another integration constant.

With the piecewise constant driving force $F(x)$, the second derivative of $P_0$ with respect to $\theta$ takes the simple form:
  \begin{equation}
    \partial^2_{\theta\theta} P_0 = \begin{cases}
        \zeta_0(\theta) F(x) & \mbox{for} \abs{\theta} < \thetaconemin
        \\
        0 & \mbox{otherwise}
    \end{cases},
\end{equation}
where we have defined 
\begin{equation*}
    \zeta_0(\theta) \equiv \frac{\bar v \left[ \bar v (\cos 2\theta-3) - 2 \sigma_F \cos\theta \right]}{4\pi \left[\bar v \cos\theta+\sigma_F \right]^3},
\end{equation*}
together with $\sigma_F \equiv \fmax + \fmin$.

Since $P_0$ already ensures normalization of the full distribution, we require $\mathcal{P}_1(\theta) = \int_0^1\dd{x} P_1(x,\theta) = 0$.
Using that $j_1(\theta)=0$ for $\abs{\theta} > \thetaconemin$
and performing the integral over $x$ for the other case, 
we find the combined solution
\begin{equation}\label{eq_j1_appendix}
    j_1(\theta) = \begin{cases}
        -\frac{1}{2}\zeta_0(\theta) & \mbox{for} \abs{\theta} < \thetaconemin
        \\
        0 & \mbox{otherwise},
    \end{cases}
\end{equation}
yielding
\begin{equation*}
    P_1(x,\theta) = \begin{cases}
        \frac{-\bar v \zeta_0(\theta)}{\bar v \cos\theta + F(x)}\left[ \frac{1}{2} + V(x)\right] & \mbox{for} \abs{\theta} < \thetaconemin
        \\
        0 & \mbox{otherwise}
    \end{cases}.
\end{equation*}

Using Eqs.~\eqref{eq_j0_appendix} and \eqref{eq_j1_appendix} we further approximate the total particle current for $\bar v \gg 1$ up to $\order{\bar v^{0}}$ as
\begin{align*}
    \current_\mathrm{nc} &= \int_{-\pi}^{\pi} \dd{\theta} \left[ j_0(\theta) + j_1(\theta) \right] \nonumber
    \\
    &= \frac{\sqrt{\bar v^2 - (1-\lambda)^{-2}}}{\pi} \left( 1 + \frac{\lambda^2}{2} \right) + \mathcal{G}(\lambda, \bar v)
\end{align*}
where
\begin{equation}
    \mathcal{G}(\lambda, \bar v) = -\frac{2}{\pi} \frac{\arctan\left[ \sqrt{\frac{\sigma_F - \bar v}{\sigma_F + \bar v} } \tan\frac{\thetaconemin}{2} \right]}{\lambda (1-\lambda)\sqrt{\sigma_F^2 - \bar v^2}},
\end{equation}
for $\bar v \in [\abs{\fmin}, \sigma_F)$, and
\begin{equation}
    \mathcal{G}(\lambda, \bar v) = -\frac{2}{\pi} \frac{\artanh\left[ \sqrt{\frac{\bar v - \sigma_F}{\bar v + \sigma_F} } \tan\frac{\thetaconemin}{2} \right]}{ \lambda (1-\lambda)\sqrt{\bar v^2 - \sigma_F^2}}.
\end{equation}
for $\bar v \in (\sigma_F, \fmax]$.\\

\renewcommand \thefigure{B\arabic{figure}}
\setcounter{figure}{0}
\setcounter{equation}{0}
\renewcommand{\theequation}{B\arabic{equation}}

\paragraph{\label{app:displacement_current}Appendix B: Current for large velocities in the presence of cyclic driving}
In this section, we describe an approximation scheme that allows us to predict the emergence of current reversal and estimate the corresponding current magnitude.
To calculate the current over a full period of the cyclic drive, 
we consider the active and passive phases individually.
Moreover, since reversals typically occur for periods $T \ll 1/D_r$, 
we neglect the influence of rotational noise and assume that the particle can self-propel only along the $\pm \hat{\vect{x}}$ directions.

When the particle's self-propulsion speed is sufficiently large to cross both sides of the potential,
we define $\tau_\pm$ as the times it takes to travel a full spatial period of the potential along $\pm \hat{\vect{x}}$.
These times thus satisfy
\begin{equation}
\pm 1 = \pm \vmax \tau_\pm + \int_0^{\tau_\pm} \dd{t} F(x(t)).
\end{equation}
Performing the change of variable $t \to x$ in the integral, we then obtain
\begin{align*}
        \pm 1 & = \pm \vmax \tau_\pm + \int_0^1 \dd{x} \frac{F(x)}{\abs{\dot{x}}}
        \\
        & = \pm \vmax \tau_\pm + \frac{\lambda \fmax}{\abs{\fmax \pm \vmax}} + \frac{(1-\lambda)\fmin}{\abs{\fmin \pm \vmax}}
\end{align*}
which, after replacing $\fmin$ and $\fmax$ by their corresponding expressions and solving for $\tau_\pm$, 
yields Eq.~\eqref{eq_taupm} of the main text.
Assuming left and right orientations to be equiprobable, we then estimate the mean particle displacement $\langle \Delta x_{\rm a} \rangle$ during the active phase by approximating the effective particle velocity as $\tau_\pm^{-1}$, such that
\begin{equation}
	\langle \Delta x_{\rm a} \rangle = \alpha T \left( \frac{1}{\tau_+} - \frac{1}{\tau_-} \right).
\end{equation}
In addition, noting that 
$$
\tau_- - \tau_+ = \frac{2(1-2\lambda)}{(1-\lambda^2 \vmax^2)(1 - (1-\lambda)^2 \vmax^2)},
$$
we conclude that for $0 \le \lambda < \tfrac{1}{2}$ and $\vmax > \fmax$,
the displacement $\langle \Delta x_{\rm a} \rangle$ remains always positive. 

To estimate the mean displacement in the passive phase, 
we assume that particles are initially uniformly distributed in space,
which matches with our observations in the limit of large self-propulsion speeds in the active phase (see \ref{fig:4} of the main text).
Particles then simply drift towards the potential minimum, 
while they may or may not reach it depending on the duration of the passive phase and their initial position.
In practice, we must consider three cases determined by the time scales 
$\Tresetleft = \tfrac{\lambda^2}{1-\alpha}$ and $\Tresetright = \tfrac{(1-\lambda)^2}{1-\alpha}$.

For $T > \Tresetright$ the particle falls to the bottom of the potential regardless of its initial position, 
such that the mean displacement reads 
\begin{equation}
\langle \Delta x_{\rm p} \rangle = \int_0^1 \dd{x} (\lambda - x) = \lambda - \frac{1}{2} \quad (T > \Tresetright).
\end{equation}
For $\Tresetleft < T \le \Tresetright$, a particle initially on the shallower side of the potential will not reach the minimum
within the duration $(1 - \alpha)T$ of the passive phase if its position $x > \xresetright$ with $\xresetright = \lambda - \fmin (1-\alpha)T$. 
We then have
$\langle \Delta x_{\rm p} \rangle = \int_0^{\xresetright} \dd{x} (\lambda - x) + (1 - \xresetright) \fmin (1-\alpha)T$,
which after some calculation leads to
\begin{equation}
\langle \Delta x_{\rm p} \rangle = \lambda - \frac{1}{2} + \frac{\left( 1 - \xresetright \right)^2}{2} \quad (\Tresetleft < T \le \Tresetright),
\end{equation}
and yields Eq.~\eqref{eq_passive_current_calc} after replacing $\xresetright$ with its expression.
Finally, when $T \le \Tresetleft$ the particle may not reach the potential minimum in a time $(1-\alpha)T$ 
if its position $x < \xresetleft$ with $\xresetleft = \lambda - \fmax (1-\alpha) T$.
With a similar calculation as above, we then obtain
\begin{equation}
\langle \Delta x_{\rm p} \rangle = \lambda - \frac{1}{2} + \frac{\left( 1 - \xresetright \right)^2 - \xresetleft^2}{2} \quad (T \le \Tresetleft).
\end{equation}

In the end, we obtain an estimation of the particle current as follows
\begin{equation}
    \current = \frac{\langle \Delta x_\mathrm{a}\rangle + \langle \Delta x_\mathrm{p} \rangle}{T},
\end{equation}
for a full period.\\

\renewcommand \thefigure{C\arabic{figure}}
\setcounter{figure}{0}
\setcounter{equation}{0}
\renewcommand{\theequation}{C\arabic{equation}}

\paragraph{\label{app:details}Appendix C: Details on numerical methods}
We discretize Eq.~\eqref{eq_Langevin} using the standard Euler-Maruyama scheme for stochastic differential equations
\begin{align}
    \vect{x}(t+\dt) &= \vect{x}(t) + \dt \left[ v(t)\unitvect{e}(\theta(t))-\mu\nabla V(\vect{x}(t)) \right]
    \\
    \theta(t+\dt) &= \theta(t) + \sqrt{2 D_r \dt} \, \eta(t).
\end{align}
Usually, we simulate until a final time of $t_f = n T/D_r$ where $n \in \mathbb{N}$ is such that $n\geq 10$ and $t_f \geq 5000/D_r$ for a given period $T$ ensuring that the system reaches steady state.
Averages are generally performed over at least 1000 trajectories.
Depending on the parameters, we choose $\dt$ such that crossing over either side of the potential when starting in the potential minimum requires at least $5$ $\dt$-steps (for $D_t=0$).
Specifically we use $\dt \leq 10^{-4}/D_r$ for the different simulations.
\\
The steady state current $\current$ is estimated from the mean displacement as
\begin{equation}
    \current \approx \frac{\mean{x(t_f)} - \mean{x(t_0)}}{t_f - t_0}
\end{equation}
with the final simulation time $t_f$ described above and some initial time $t_0=T$ for the cyclic simulations.
\end{document}